\documentclass{article}
\usepackage{latexsym, amsmath, amssymb, enumerate, graphics}

\newtheorem{theorem}{Theorem}
\newtheorem{lemma}{Lemma}

\newcommand{\prfe}{\hfill $\Box$ 

                                                  \smallskip}

\numberwithin{equation}{section} 
\title{On Blowup for Gain-Term-Only classical and relativistic Boltzmann equations}
\date{}
\author{\sc{
H\aa{}kan Andr\'{e}asson}\footnote{Department of Mathematics, Chalmers University of Technology, S-412 96 G\"{o}teborg, Sweden},\\ 
\sc{Simone Calogero}\footnote{Max-Planck-Institut f\"ur Gravitationsphysik, Am M\"uhlenberg 1, D-14476 Golm, Germany},\\ 
\sc{Reinhard Illner}\footnote{Department of Mathematics and Statistics, University of Victoria, Victoria, B.C. V8W 3P4, Canada}}

\begin{document}
\maketitle
\begin{abstract}
  We show that deletion of the loss part of the collision term in all physically relevant 
  versions of the Boltzmann equation, including the relativistic case, will in general lead 
  to blowup in finite time of a solution and hence prevent global existence. Our result corrects 
  an error in the proof given in Ref. \cite{IS2}, where the result was announced for the 
  classical hard sphere case; here we give a simpler proof  which applies much more generally.
 \end{abstract} 
{\bf Key words. } Boltzmann equation, relativistic Boltzmann equation, blowup\\ 
\phantom{hej}\\ {\bf AMS subject
  classifications. }76P05, 83A05, 35S30

\section{Introduction}\label{intro}
It has been known for some time that the classical Boltzmann equation admits global unique 
solutions in all space if either the initial data are small enough in suitable norms, or,
equivalently, the mean free path is large enough. The first result of this type appeared 
in \cite{IS}, where the analysis was restricted to the hard sphere case. Hamdache
\cite{H} generalized the method to cut-off hard potential interactions, and Bellomo and Toscani 
found a variation of the method which allowed inclusion of less rapidly decreasing initial values \cite{BT}.
The method and the key references may be found in research monographs such as \cite{CIP, G} or in the survey article by Ukai \cite{U}. 
The regularity of these solutions has more recently been explored in work by Mischler and Perthame \cite{MP}, 
Boudin and Desvillettes \cite{BD}.

The key property which made these global existence and uniqueness results  possible is 
the hidden convolution structure present in the gain part of the collision term; this structure
was first observed, and exploited in the discrete velocity setting in one dimension by Tartar \cite{T}. 

Casual analysis of the quoted literature shows that the method does actually barely
use the decomposition of the Boltzmann collision operator into a gain and a loss part; in fact,
the results remain true for gain-term-only Boltzmann equations, i.e., equations where the loss
part is simply deleted from the collision term. However, no such
global result has been proved for the relativistic Boltzmann
equation. Here the collision geometry is more involved (see the end
of this section) and the available methods of proof 
in the classical case depend 
crucially on the collision geometry, eg. the method of 
Illner and Shinbrot relies on the identity (3.5) in \cite{IS}, which
is a consequence of the Galilean invariance of the classical collision 
geometry. (See also the discussion in Section 2.5 of \cite{G}). 

We remark that it has been known for a long time that in each 
case where the gain-term-only Boltzmann equation can be solved uniquely on some time interval, 
then the 
full Boltzmann equation can be solved, uniquely, on the same time interval for the same initial
value. A good method to prove this is the Kaniel-Shinbrot iteration scheme (see \cite{KS}). 
The methods in \cite{IS} and the other mentioned references really first focus
on gain-term only equations. Once these are solved, it is then a rather mechanical matter to 
include the loss terms.

The methods will certainly not work for spatially homogeneous gain-term-only Boltzmann 
equations, because (unless
one considers zero data) the spatial homogeneity implies that the total amount of gas present is 
infinite.  One of the results we present below is a simple argument which shows that we will 
in general have blowup for the homogeneous case in finite time even if velocity space is 
artificially truncated to a bounded domain. 

Here, we demonstrate the limits of the method by showing that deletion of the loss term will 
in general mean loss of global solvability for all relevant space-dependent 
versions of the Boltzmann equation,
including the relativistic case. Such a result was proved in \cite{IS2} for 
classical hard spheres, but, as recently observed by one of the authors (H. A.), the 
proof given there is flawed. Specifically, the method in \cite{IS2} uses integration
of suitably chosen subsets of velocity space to obtain lower bounds for solutions. Near the bottom
of page 256 of the reference, a set ${\cal M}_>^+$ is defined to complete a certain estimate. 
A careful geometric analysis shows that this set is always empty, and the argument  
collapses at this point. R. Illner, co-author of the reference, regrets this error, 
but is pleased to present a better method in this paper. 

The key steps presented in Section \ref{classical} below depend mostly on the collision geometry, which is 
quite simple in the classical cases: Post- (or pre-) collisional velocities lie opposite on a 
sphere with center $1/2 (v+w)$ and diameter $\|v-w\|$, where $v$ and $w$ are the velocities of 
the collision pair under consideration. 
For relativistic interactions the collision geometry is different from the
classical one, due to the fact that the collison invariants have a
different form. The collision geometry for
classical interactions is spherical and invariant under translations
(Galilean invariant), 
i.e., the collision sphere is unchanged as long as the relative
velocity remains the same.  This is not the case in the relativistic
situation where the collision geometry is Lorentz invariant; instead
of spheres we get ellipsoids and the excentricity of an
ellipsoid changes with the energies of the particles involved in the
collision process. The collision geometry is 
central in Lemma 1 below but the properties of the collision geometry
needed in the proof are easily verified in both cases.

\section{The Gain-Term-Only Boltzmann Equation}\label{classical}
We are concerned with the equation
\begin{equation}
\partial_tf+v\cdot\nabla_xf=Q_+(f,f)
\end{equation}
with initial condition
\begin{equation}
f(0,x,v)=\varphi(x,v).
\end{equation}
Here, $t\ge0$ denotes time, $\varphi$ is an integrable nonnegative initial value, and 
\begin{equation}\label{gain}
Q_+(f,f)(v)= \int_{{\Re}^3} \int_{S_+^2}
B(|v-w|,n\cdot(v-w))f(v')f(w')\,{\rm d}n {\rm d}w, 
\end{equation}
denotes the gain part of the usual collision term for the Boltzmann equation, in standard notation 
(the local dependence on $t$ and $x$ in (\ref{gain}) is omitted). Pre- and post- collisional velocities satisfy the conservation
of momentum, $v+w=v'+w'$, and of energy, $|v|^2+|w|^2=|v'|^2+|w'|^2$.
$B$ is the collision kernel which depends on the type of interparticle potential. For 
details, see \cite{C}. In the hard sphere case, considered in \cite{IS2}, 
$B=n\cdot(v-w).$  For this case it is shown there that the initial value problem has a 
local unique solution; the point is that this solution will in general blow up in finite time.

Our present objective is to provide a correct and more general proof for this fact. The proof will
work for all kernels $B$ satisfying the condition that if $M\in
S^2_+\times\Re^3$ is a set of positive measure, then 
\begin{equation}\label{assumption}
\int\int_M B\, {\rm d}n{\rm d}w>0. 
\end{equation} 
This assumption is used in the proof of Lemma \ref{char} below and does not seem to exclude anything
physically reasonable.
We will further work under the assumption that the initial 
value problem is uniquely solvable locally in time, a fact which can be proved for reasonable
collision kernels along the lines of the proof in \cite{KS}. 

As in \cite{IS2} we restrict our analysis to initial values 
$\varphi(x,v)=c_0 \chi_1(x)\chi_2(v)$ where $\chi_1$ and $\chi_2$ are characteristic functions 
of the balls $B_1=\{ x; |x|\le c_1\}, B_2=\{v;|v|\le c_2\}$, and $c_1,c_2$ are positive
constants to be chosen later.
 
Before we can deal with the full problem, we focus on the spatially homogeneous case with a 
truncation in velocity space. The reason for the truncation will become transparent later.

A key ingredient in our argument is the following estimate for the
action of the gain term on characteristics functions.\footnote{We are grateful to
Cedric Villani for pointing out this estimate to us.} 

\begin{figure}[htbp]
\begin{center}\scalebox{.5}{\includegraphics{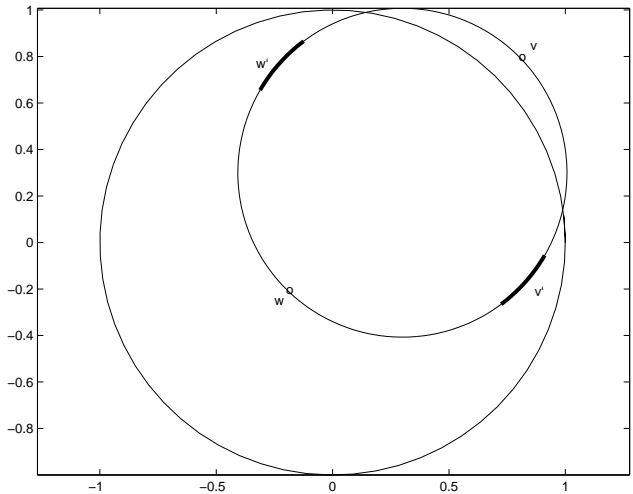}}
\end{center}
\caption{Non-relativistic collision geometry}\label{fig1}
\end{figure}

\begin{lemma}\label{char}
Let $B_R=\{v; |v|\le R \}$ and assume that $B$ satisfies (\ref{assumption}). 
Then there are constants $\lambda>1$ and $\delta>0$ such that 
\begin{equation}
Q_+(\rho \chi_{B_R},\rho \chi_{B_R})(v)\ge \delta \rho^2 \chi_{\lambda B_R}(v) \ge 
\delta \rho^2 \chi_{B_R}(v). 
\end{equation}
\end{lemma}
{\it Proof:}  
This estimate is a consequence of the collision geometry, see Figure
\ref{fig1} below (one dimension is suppressed).\footnote{Figures have been obtained using the software Matlab.} Choosing $v$ in 
or just  outside the ball $B_R$, we can construct collision spheres such that both the centre of 
the sphere and the antipode $w$ of $v$  are both in $B_R$, and  the post-collisional velocities 
$v',w'$ are also in $B_R$ for a set of collision parameters $n\in S^2_+$ of positive measure on $S^2$.
The constant $\delta$ is a positive lower bound of  $I_v:=c{\int\int}_M B(|v-w|,n\cdot(v-w)) {\rm d}n{\rm d}w$, 
where $M$ is the positive measure set of $w\in B_R$ and $n\in S^2_+$ for which the above conditions 
hold. It is clear that  $I_v$ has a positive lower bound unless $\lambda$ is chosen too large or unless 
$B$ vanishes on sets of positive measure, which is prohibited by the assumption (\ref{assumption}). \prfe

The proof of the next lemma is immediate, but we formulate the result for easy reference.

\begin{lemma}
Let $f\ge g \ge 0$ be integrable density functions for which $Q_+$ is defined. Then 
$Q_+(f,f)(v)\ge Q_+(g,g)(v).$
\end{lemma}

Next, we define  $Q_R(f,f)(v)= \inf_v \{\chi_{\{|v|\le R\}} Q_+(f,f)(v)\}.$
Note that $Q_R$ is defined such that $Q_R(v)=0$ if $|v|>R$, and that $Q_R(v)$ takes a constant 
value for $|v|\le R.$  

Consider an initial value $f_0=f_0(|v|)$  (the isotropy in $v$ is not essential but convenient)
such that there is a constant $\rho_0>0$ with  $f_0(|v|)\ge \rho_0 \chi_{B_R} (v).$ Let 
$f(t,v)$ be the (local) nonnegative solution of 
\begin{equation}
\frac{d}{dt}f=Q_+(f,f),  \ \  f(0,v)=f_0(v)
\end{equation}
and let $\tilde{f}(t,v)$ be the solutions of 
\begin{equation}
\frac{d}{dt}\tilde{f}=Q_R(\tilde{f},\tilde{f}),  \ \  \tilde{f}(0,v)=\rho_0 \chi_{B_R}(v). \label{*}
 \end{equation}

\begin{lemma}
\[ f(t,v)\ge \tilde{f}(t,v)   \]
while $f(t,v)$ exists.
\end{lemma} 
{\it Proof:} Let $[g-h]_+$ equal $g-h$ if $g>h$ and $0$
otherwise. We derive a Gronvall inequality for
$\|[\tilde{f}(t,\cdot)-f(t,\cdot)]_+\|_{\infty}$ which shows that it 
remains zero as long as the solution exists. 

We have 
\begin{eqnarray}
\|[\tilde{f}(t,\cdot)-f(t,\cdot)]_+\|_{\infty}&=&\|[\tilde{f}_0(v)-f_0(v)+\int_{0}^t
Q_R(\tilde{f},\tilde{f})-Q_+(f,f){\rm d}s]_+\|_{\infty}\nonumber \\
&\leq&\int_{0}^t
\|[Q_+(\tilde{f},\tilde{f})-Q_+(f,f)]_+\|_{\infty} {\rm d}s,\label{>1}
\end{eqnarray}
by the definition of $Q_R.$ 
The integrand in (\ref{>1}) can be written and estimated as follows 
\begin{eqnarray}
&\|[\int\int
B\tilde{f}(v')(\tilde{f}(w')-f(w'))+
Bf(w')(\tilde{f}(v')-f(v')){\rm d}n{\rm
  d}w]_+\|_{\infty}&\nonumber \\
&\leq C\|[\tilde{f}(t,\cdot)-f(t,\cdot)]_+\|_{\infty},& 
\end{eqnarray}
where $$C=\|\int\int
B\tilde{f}(v'){\rm d}n{\rm d}w\|_{\infty}+\|\int\int
Bf(w'){\rm d}n{\rm d}w\|_{\infty}.$$ 
Since $C$ is finite on the interval of existence the claim follows
from (\ref{>1}). \prfe  

\begin{lemma} 
The solution of (\ref{*})  is of the form  $\tilde{\rho}(t)\chi_{B_R}(v)$.
\end{lemma}
{\it Proof:} This follows because $\tilde{f}(0,v)=\rho_0 \chi_{B_R}(v)$ and because 
$Q_R(\tilde{f},\tilde{f})$ is homogeneous on $B_R$ by construction, and zero elsewhere.\prfe

\begin{theorem}\label{blowup}
$\tilde{f},$ and hence $f$, both blow up in finite time.
\end{theorem}
{\it Proof:}  For any $v\in B_R$,  
\begin{eqnarray}
\frac{d}{dt}\tilde{f}(v) 
&=& 
\frac{d}{dt}\tilde{\rho}(t) \nonumber \\
&=& 
Q_R(\tilde{f},\tilde{f}) \nonumber \\
&\ge& 
\inf_{|v|\le R} Q_+(\tilde{\rho} \chi_{B_R}, \tilde{\rho} \chi_{B_R}) 
\nonumber \\
&\ge& 
\delta \tilde{\rho}^2(t). \nonumber 
\end{eqnarray}
Hence  $\frac{d}{dt}\tilde{\rho} \ge \delta \tilde{\rho}^2,$  and the assertion follows.\prfe

We return now to the spatially inhomogeneous case.
The rest of our argument is identical to what was done on \cite{IS2}, but we repeat the 
analysis here for the convenience of the reader. First, recall that we are concerned with
\begin{equation}
\partial_tf+v\cdot\nabla_xf=Q_+(f,f) \label{BE},
\end{equation}
with initial values $\varphi(x,v)=c_0 \chi_1(x)\chi_2(v).$ It is known \cite{KS} that the 
initial value problem has a unique nonnegative mild local solution under reasonable assumptions 
on the collision kernel $B$. Moreover, if we consider the modified problem 
\begin{equation}
\partial_t\hat{f}+v\cdot\nabla_x\hat{f}=\chi_2(v) Q_+(\hat{f},\hat{f}), \ \hat{f}(0,x,v)=\varphi(x,v),
\label{TBE}
\end{equation}
then this problem is also uniquely solvable in the mild sense, and  $f \ge \hat{f} $ while the larger 
of the two exists. To show blowup it is therefore enough to show that $\hat{f}$ will blow up in finite 
time, and to this end we will reduce the problem to the spatially homogeneous case resolved in the 
previous section. We need 
\begin{lemma}
Choose $T>0$ arbitrary but fixed, and suppose that $c_1>Tc_2.$ Then, if $0\le t <T$ is in the interval
of existence of $\hat f$, and if $|x|,|y|\le c_1-t c_2,$ then $\hat{f}(t,x,v)=\hat{f}(t,y,v)$ for all
$v.$ 
\end{lemma}
{\it Proof:} This proof is identical to the one in \cite{IS2}. First, note that the local solution
of the initial value problem is obtained as the limit as $k\rightarrow \infty$ of the iteratively 
defined functions $\hat{f}_k,$ where   $\hat{f}_0=0$ and 
\begin{eqnarray} 
\lefteqn{\hat{f}_{k+1}(t,x,v)=c_0 \chi_1(x-tv)\chi_2(v)} \nonumber \\
&+\chi_2(v)\int_0^t \int \int B\hat{f}_k(\tau,x-(t-\tau)v,v') 
 \hat{f}_k(\tau,x-(t-\tau)v,w')\,{\rm d}n {\rm d}w.  \label{iterate} 
\end{eqnarray}
We also note that $\hat{f}_{k+1}\ge \hat{f}_k$, and by construction  $f_k(v)=0$ for all $k$ and 
all $|v|> c_2.$  The claim of the lemma is trivial for $k=0,$ and assuming that 
it holds for some $k>0,$ it follows from (\ref{iterate}) that it also holds for $k+1$: in fact,
if one of $|v'|$ or $|w'|$ is larger than $c_2,$ the integrand in (\ref{iterate}) vanishes. If
$\max\{|v'|,|w'|\}\le c_2,$ and $|x|\le c_1-t c_2,$ we have in addition that  
$|x-(t-\tau)v|\le |x|+(t-\tau)c_2 \le c_1-\tau c_2. $  The inductive hypothesis then entails that 
\[\hat{f}_k(t,x-(t-\tau)v,\zeta)=\hat{f}_k(t,y-(t-\tau)v,\zeta)\]
for both $\zeta=v', \zeta=w'$ whenever $|x|,|y|\le c_1-t c_2.$ 
It follows that for all such $x,y$, all $v$ and all $t\in [0,T]$ that 
 $\hat{f}_{k+1}(t,x,v)=\hat{f}_{k+1}(t,y,v).$\prfe
 
The next Theorem is our main result.

\begin{theorem}\label{blowup2}
Let $f$ be the local mild solution of (\ref{BE}) with the initial value 
$\varphi(x,v)=c_0 \chi_1(x)\chi_2(v).$ If $c_0$ or $c_1$ are sufficiently large this solution 
will blow up in finite time, in both the $L^1-$ or the $L^{\infty}-$ norm.
\end{theorem}
{\it Proof:}  If we consider a fixed time interval $[0,T]$ and a fixed $c_2>0,$ let $c_1>T c_2.$
Then the previous lemma implies that $\hat{f}$ is independent of $x$ in the set 
$\{ (t,x); 0\le t\le T, |x|\le c_1-t c_2 \}$, so on this set the evolution of $\hat{f}$ is effectively
given as the solution of the initial value problem
\begin{equation}
\frac{d}{dt} \hat{f} (v) = \chi_2(v) Q_+(\hat{f},\hat{f}), \  \hat{f}(0,v)=c_0 \chi_2(v).
\end{equation}
We know from our earlier discussion of the truncated spatially homogeneous case that the solution
of this initial value problem will blow up in $[0,T]$ if $c_0$ is large enough; alternatively, 
if $c_0$ is positive but small, we can make $c_1$ large enough such that the condition 
$c_1> c_2 T$ will hold for the required blowup time $T$. The assertion of the theorem follows.\prfe


\section{The relativistic case}\label{relativistic}
This section contains a discussion on  the generalization of Theorem \ref{blowup2} to the relativistic Boltzmann
equation. We start with a short introduction to the problem, see \cite{CK} for more details.

The speed of light and the rest mass of particles
are both normalized to unity, i.e., $m=c=1$; the convention on the signature is $(+ - - - )$. We denote by $f=f(t,x,p)$ the
one-particle distribution function, with $p\in\Re^3$ denoting the momentum variable. The relativistic Boltzmann equation has the form
\begin{equation}
(\partial _{t} + \frac{p}{p_{0}}\cdot\nabla _{x})f=Q(f,f), \label{rbe}
\end{equation}
where the collision operator is defined by
\begin{equation}
Q(f,g)(p)=\frac{1}{p_{0}}\int_{\Re^{3}}\int_{{S}^{2}}(f(p')g(q')-f(p)g(q))B(g,\theta)
{\rm d}\Omega\frac{{\rm d}q}{q_{0}}.\label{col}
\end{equation}
Here ${\rm d}\Omega$ is the element of surface area on ${S}^{2}$,
$p^{\mu}=(p_{0},p)$ is the four momentum ($p_{0}\in{\Re}^+,\;
p\in{\Re}^{3},\;\mu =0,1,2,3$), and $p_{0}=\sqrt{1+|p|^{2}}$ is the
particle energy. Conservation of momentum and energy now read
\[
p+q=p'+q',\quad \sqrt{1+|p|^2}+\sqrt{1+|q|^2}=\sqrt{1+|p'|^2}+\sqrt{1+|q'|^2}.
\]
The total energy and the relative momentum in the
center of mass system are $s^{1/2}=\mid q^{\mu}+p^{\mu}\mid$ and
$2g=\mid q^{\mu}-p^{\mu}\mid$, respectively. The scattering angle $\theta$ in
the center of mass system satisfies
\begin{equation}
\cos\theta=1-2\frac{(p^{\mu}-q^{\mu})(p'_{\mu}-q'_{\mu})}{|p^{\mu}-q^{\mu}|^{2}}.
\end{equation}
The kernel $B(g,\theta)$ and the scattering cross section $\sigma(g,\theta)$ are related
by
\begin{equation}
B(g,\theta)=\frac{gs^{1/2}}{2}\sigma(g,\theta).
\end{equation}
The case of a constant scattering cross section corresponds to the relativistic hard sphere model. For Maxwellian molecules one has
$\sigma=g^{-1}(1+g^2)^{1/2} F(\theta)$, where $F$ is an arbitrary function of the scattering angle.

We also mention a different representation of the relativistic
collision operator which is closer in form to the classical one, see 
Appendix II in \cite{GS2} for a derivation. We can write
\begin{equation}
Q(f,g)=\int_{{\Re}^{3}}\int_{{S}^{2}} k(p,q,\omega)[f(p') g(q')-f(p)g(q)]\,{\rm d}\omega{\rm d}p 
\end{equation}
where $p'=p+a(p,q,\omega)\omega,\,q'=q-a(p,q,\omega)\omega$ and 
\[
k(p,q,\omega)=4s\sigma (p_{0}+q_{0})^{2}\frac{|\omega\cdot (\hat{q}-\hat{p})|}{(e^{2}-(\omega\cdot (p+q)^{2}))^{2}},
\]
\[
a(p,q,\omega)=\frac{2ep_{0}q_{0}(\omega\cdot(\hat{q}-\hat{p}))}{e^{2}-(\omega\cdot(p+q))^{2}}.
\]
Here $e:=p_{0}+q_{0}$ is the total energy and
$\hat{x}:=x/x_{0}$.
As shown in Figure \ref{fig2} below, for fixed $p,q$ and as $\omega$ varies on $S^2$, $p'$ and $q'$ lie on an ellipsoid (one
dimension is suppressed). The excentricity of the ellipsoid is given by 
\[
\alpha=\frac{\sqrt{1+|p|^2}\sqrt{1+|q|^2}}{\sqrt{1+|p'|^2}\sqrt{1+|q'|^2}}.
\]  
In particular it is bounded for bounded pre-collisional momenta. 

Finally, the collision operator can in an obvious way be written
as $$Q(f,g)=Q_{+}(f,g)-Q_{-}(f,g),$$ where $Q_{+}$ and $Q_{-}$ are
referred to as the gain and loss term respectively. Hence, the
gain-term-only relativistic Boltzamnn equation reads, 
\begin{equation}
(\partial _{t} + \hat{p}\cdot\nabla _{x})f=Q_+(f,f)\label{RB2}.
\end{equation}

Local existence and uniqueness of mild solutions to the Cauchy problem for the relativistic Boltzmann equation has been shown in \cite{B},
where the more general case of a background curved metric is
considered. For initial data close to equilibrium global existence has 
been obtained in various function spaces \cite{GS1, GS2}.  At present
a similar result for data near vacuum, i.e., for small data, is
not available as was briefly discussed in the introduction. 

By inspection of the proof of Theorem 1 it follows that an identical
result also holds for the gain-term-only relativistic Boltzmann
equation (\ref{RB2}). The essential properties of the collision
geometry needed in the proof of Lemma 1 are easily verified also in
this case. The relevant picture in the relativistic case is given by
Figure \ref{fig2} below. 
  
\begin{figure}[htbp]
\begin{center}\scalebox{.5}{\includegraphics{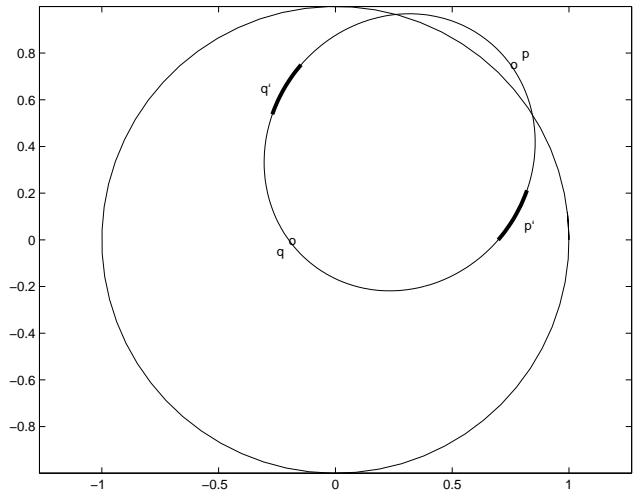}}
\end{center}
\caption{Relativistic collision geometry}\label{fig2}
\end{figure}

  
The remaining details of the proof of blowup for the only-gain-term relativistic Boltzmann equation are very similar to the classical case and are omitted. Note however, that since finite propagation speed is already present in the relativistic situation, velocity truncation is not longer necessary in this case.  

\noindent{\bf Acknowledgment:}
S.~C.\ acknowledges support by the European HYKE network 
(contract HPRN-CT-2002-00282).

\end{document}